\documentclass[twocolumn,showpacs,preprintnumbers,amsmath]{revtex4}
\usepackage{epsbox}
\usepackage{graphicx}% Include figure files
\usepackage{dcolumn}% Align table columns on decimal point
\usepackage{bm}% bold math

\begin{document}

\preprint{APS/123-QED}

\title{Inevitable Irreversibility Generated by the Glass Transition \\
of the Binary Lattice Gas Model
}

\author{Miki Matsuo}
%\email{miki@jiro.c.u-tokyo.ac.jp}
\affiliation{
Department of Pure and Applied Sciences, University of Tokyo. \\
Komaba, Meguro-ku, Tokyo 153
}%Lines break automatically or can be forced with \\

\date{\today}

\begin{abstract}
We numerically investigate the thermodynamic properties of the glass state.
As the object of our study,
we employ a binary lattice gas model.
Through Monte Carlo simulations, we find that this model actually 
experiences a glass transition.
We introduce a potential into the model that 
represents a piston with which we
compress the glass.
By measuring the work performed in this process,
we find that irreversible works exist at the glass state
even in the quasistatic limit.
This implies that yield stress is created by the glass transition.
\end{abstract}

\maketitle

\begin{figure}
\begin{center}
\epsfile{file=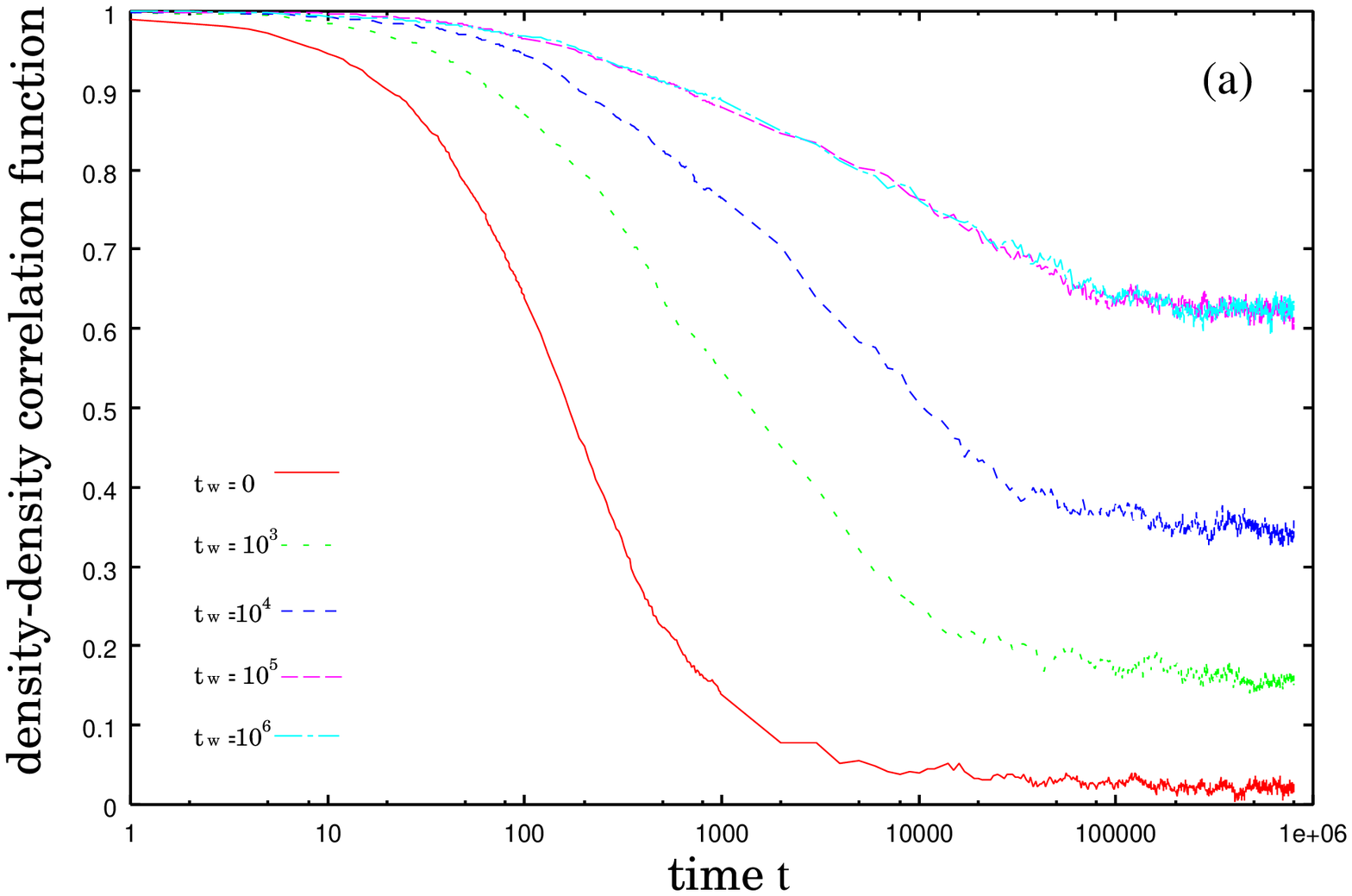,scale=0.4}
\epsfile{file=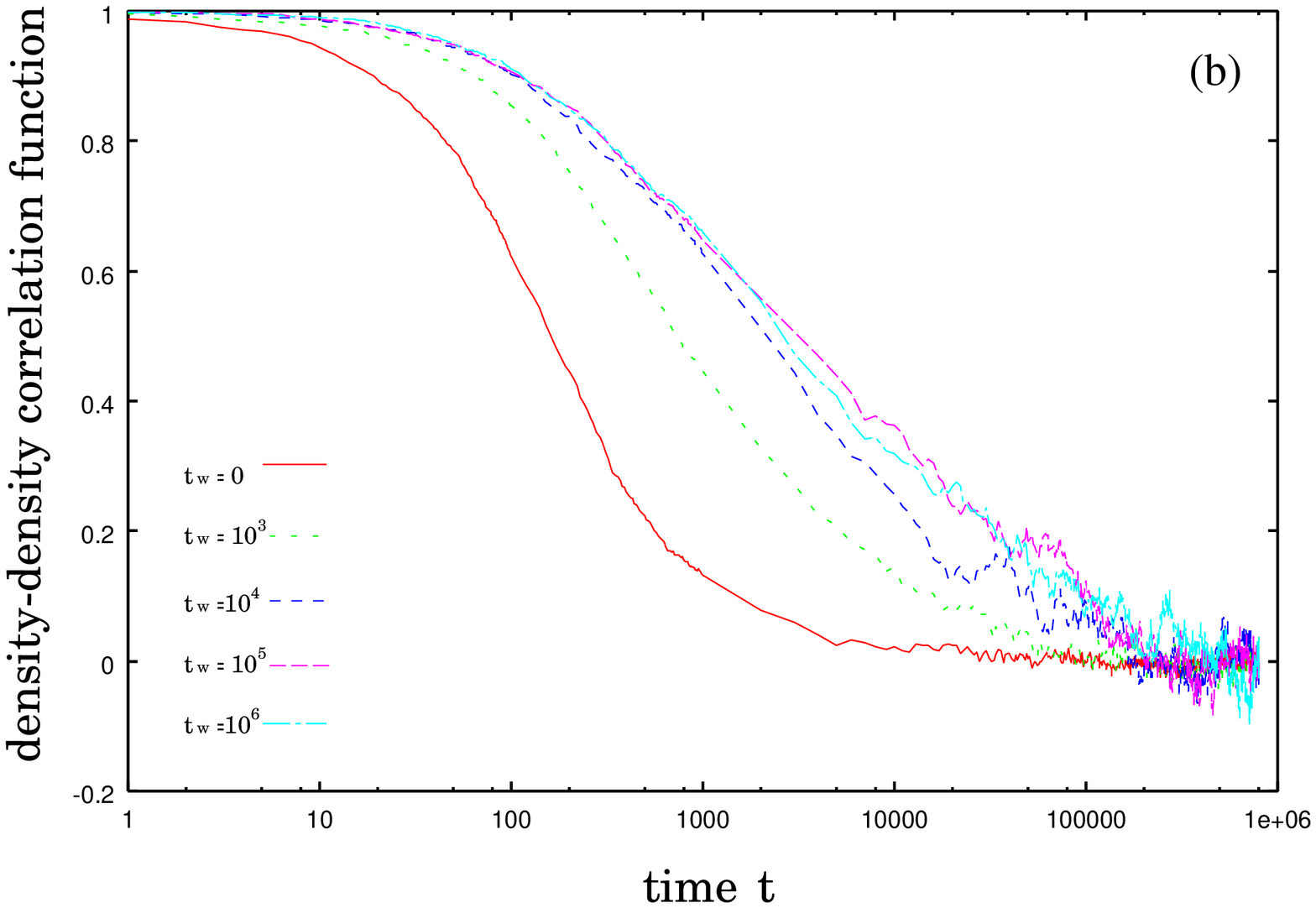,scale=0.4}
\end{center}
\caption{Monte Carlo simulation results for 
the density-density correlation function $\phi(t+t_w,t_w)$ 
of the binary lattice gas.
Correlations are plotted for several waiting times $t_w$:
$t_w=0,10^3,10^4,10^5,10^6$.
(a) The parameter values are here $(N^A,N^B,T)=(65,65,1000)$. 
The unique curve in the long waiting time limit has a finite 
Debye-Waller factor $f_D$ defined by 
$f_D=\lim_{t \rightarrow \infty} \lim_{t_w \rightarrow \infty} \phi$.
The nature of the correlation dynamics in this case indicates that
the system is in the glass state.
(b) The parameter values are here $(N^A,N^B,T)=(55,55,1000)$. 
In this case, the unique curve in the 
long waiting time limit has a vanishing Debye-Waller factor.
The nature of the correlation dynamics in this case indicates that 
the system is in the gas state.}
\end{figure}

Theoretically describing structural glass is a long-standing problem
in statistical physics.
As the temperature of a supercooled liquid is reduced,
its molecular dynamics become slower,
and eventually such motion ceases at a temperature $T_g$,
which is called the glass transition temperature~\cite{EDI,STI,ANG,DIS}.
Below this temperature,
the glass is usually considered to be a solid
in the sense that its molecules are trapped within a cage
formed by their neighbors
from which 
they cannot escape,
at least within the experimental time scale.
However, a glass is never a solid 
in the true sense, but out of equilibrium.

%The most comprehensive theory is mode-coupling theory (MCT),
%proposed by Bengtzelius, G\"{o}tze and Sj\"{o}lander~\cite{GOTZE},
%and the theory succeeded to explain the universal form of the relaxation processes
%nearby $T_g$.
%Recent computer simulations in terms of 
%microscopic point of view like molecular dynamics
%revealed the complex phase space structure called inherent structure~\cite{SCI,KOB}.
%This revived classical glass theories
%as Addam-Gibbs, Kautzmann and Goldstein~\cite{ADAM,JAC}.
%Parisi et. al. have been tried to modernly form the picture
%with the replica method~\cite{Parisi,MEZ}.
%The mean field treatments including the cage effect 
%were also performed in order to
%gain another insight into the glass transition~\cite{Bou,Rhe}.

Significant effort has been devoted to understanding
the nature of glasses,
with theoretical studies employing mode coupling theory~\cite{GOTZE},
the replica model~\cite{Parisi,MEZ},
and the trap models~\cite{Bou,Rhe}.
However, in spite of these attempts,
a unified picture and satisfactory understanding have not yet been obtained.
The main source of the difficulty 
in obtaining a theoretical description is that 
we cannot use the universal measure (i.e. the canonical measure) 
in treating the glass state.
Since the glass state is not an equilibrium state,
the Boltzmann-Gibbs ensemble does not apply. 
Without the framework of statistical mechanics,
we are powerless to
theoretically treat a large number of degrees of freedom.
To allow for the treatment  of glass forming materials,
it is thus necessary to construct an extended form of 
statistical mechanics.
With this goal, as a first step,
it is wise to first consider the thermodynamics of glasses.

Since a glass represents a nonequilibrium state,
it is not definite whether thermodynamic structure
exists in this state.
%The fictive variable approach is an old method 
%which was used in glass technology,
%where the thermodynamics is assumed to approximately hold 
%including a pseudo thermodynamic variables 
%called fictive variables~\cite{Fred}.
%Recently Nieuwenhuizen proposed the glass thermodynamics,
%where the variable complexity (configurational entropy) and 
%the conjugate effective temperature
%are selected as fictive variables~\cite{Niew}.
%In his work, thermodynamic relations are modified to the new forms 
%including these pseudo variables, although
%the rough frame of classical thermodynamics remains as it is.
%However, we note that
%fictive variables are the variables which we cannot control.
%Thermodynamics should be a phenomenological low,
%that is, the theory should include only measurable parameters.
%Therefore it is doubtful the theory should be called a macroscopic law
%of the glass state.
%On the other hand, if we have no macroscopic law of the glass,
%we cannot know any essential property of the glass transition,
%because we can not even simply interpret 
%all measured quantities of the glass experiments.
For this reason we first reconsider the thermodynamic concept of a glass state.
After confirming the important macroscopic nature,
we hope to connect the macroscopic picture to the microscopic picture.
In this letter, we report the numerical investigation of the thermodynamic properties
of the glass state in a lattice gas model.

Lattice gas models are very simple and easy to treat computationally.
For this reason, if a lattice gas model possesses a glass state, 
it would be convenient for the theoretical study of this state. 
Several lattice gas models have been proposed
as models of the glass transition.
The most commonly used models are the lattice gas models with kinetic constraints
proposed by J\"{a}ckel, Palmer~\cite{DIS}, Kob and Andersen~\cite{Kob}, and Kurchan 
et al., among others.
%The hard-square lattice gas
%is proposed by J\"{a}ckel et. al,
%the hierarchcally constrained  kinetic Ising model
%is proposed by Palmer et. al. 
%More general type lattice gas is investigated by 
%Kob, Andersen and Kurchan et al.
%However, each models have some troubleness. 
%The hard-square lattice gas, which is belongs to 
%the one component lattice-gas class,
%shows a slowing down dynamics and, however,
%has no criticality.
%The hierarchcally constrained kinetic Ising model
%has criticality.
However, because these models
possess no energy, then are not suited to our thermodynamic study.
Despite this fact, we are interested in lattice-gas models because
Fuchizaki and Kawasaki assert that 
%the dynamical density functional theory (DDFT) formalism 
%directly produces 
such a  model
with a repulsive interaction
displays glassy dynamics~\cite{KAWA,KAWA2}.
Since a lattice gas model of this type does possess an energy, 
it would seem to be a possible model with which to carry out our study.
However, in our simulations, 
we have not found a glass transition in this model.
(We give detailed discussion regarding this point below.)
Therefore, we must make a new lattice model with energetic constraints.
In this letter, we extend the lattice gas model
to a two component model.
Using Monte Carlo simulations,
we find that our extended lattice gas model 
indeed exhibit a glass transition. 
Finally, we carry out  thermodynamic operations within the model and
investigate its macroscopic properties so revealed.

We first introduce our extended lattice gas model.
We consider a system defined on a 2-dimensional lattice of size $L \times L$.
We assume that there are two types of particle, A and B, on the lattice,
and that  all particles interact repulsively.
The strengths of the interactions differ
for different pairs, A and A, B and B, and A and B.
The interactions are restricted to nearest neighbors,
and there is no more than one particle at each site.
These conditions are satisfied by the Hamiltonian
\begin{equation}
{\cal H}_0 = -J^{ AA} \sum_{\langle i j \rangle} n_i^{ A} n_j^{ A} - J^{ BB} \sum_{\langle i j \rangle} n_i^{ B} n_j^{ B} - J^{ AB} \sum_{\langle i j \rangle} n_i^{ A} n_j^{ B},
\label{Hamil}
\end{equation}
where ${ n}^A_i = 1$ and ${ n}^B_i = 0$ if site $i$ is occupied by a particle
of type A, while  ${ n}^A_i = 0$ and ${ n}^B_i = 1$
if it is occupied by a particle of type B,
with $i=1,2,\cdots L \times L$ labeling
the $L \times L$ lattice points.
Here, $\langle i j \rangle$ denotes nearest neighbor pairs.
The quantities $J^{ AA}, J^{ BB}, J^{ AB}$ are the coupling constants 
between A and A, B and B, and A and B.
We call this model the "binary lattice gas model".
The model of Fuchizaki and Kawasaki
is single component version of this model.
We stipulate that the dynamics of the system are conservative
by enforcing the conditions
\begin{equation}
\sum_i n_i^A = N^A , \sum_i n_i^B = N^B.
\end{equation}
To enforce these conditions,
we adopted the Kawasaki (spin-exchange) dynamics
for the numerical simulations~\cite{Mon}.
Thus, the system is parametrized by the set of quantities 
$J^{AA},J^{BB},J^{AB},N^A,N^B,T$ and $L$,
where $T$ is the temperature of the system.
In our simulations,
negative coupling constants were used
in order to preclude the gas-liquid transition.
We set $(J^{AA},J^{BB},J^{AB})=(-1,-4,-2)$ and $L=15$,
and employed periodic boundary conditions.
The results of our Monte Carlo simulations
indicate that this model possesses a glass phase
when the particle concentration is sufficiently high.
%The most striking feature of the glass state is
%the waiting time dependence of the two time quantities like correlations. 
%That is, the two time quantities depend explicitly on the observation time 
%and on the time when the perturbation was applied.
%This kind of behavior is known as physical aging.
%In this letter, by recognizing the aging dynamics,
%we specify the glass transition.
In Fig.1 (a) we show the results for the 
density-density correlation functions with several waiting times
for the case $T=1000$ and $N^A, N^B=65$.
We see that the correlation functions exhibit an aging dynamics,
that is, the correlation depends explicitly on the observation time 
and on the time at which the perturbation is applied.
We note that when the waiting time is sufficiently long,
the correlation function converges to a universal form,
which has a finite Debye-Waller factor.
Thus we conclude that the system is in the glass state in this case with $N^A, N^B=65$.
In Fig.1 (b),
we show the density-density correlation functions for the case $T=1000$
and $N^A, N^B=55$.
We see that again in this case,
the correlation function converges to a universal form,
but here it has a vanishing Debye-Waller factor,
so that the system is concluded to be 
in the gas state in this case with $N^A, N^B=55$.
We can hence conclude that the transition between the gas state and the glass state
occurs at a concentration in the range $C = 0.42 - 0.51 $.
%The most often used parameter to judge the glass transition
%is a density-density correlation function
%\begin{equation} 
%\phi_q(t) = \langle \Delta n_q(t) \Delta n_q(0) \rangle,
%\end{equation}
%where $\Delta n_q$ denotes the fluctuation of $n_q$,
%$n_q$ denotes a Fourier component of the density field $n_i$,
%and $n_i$ is defined by $n_i \equiv n_i^A + n_i^B$.
%The idealized MCT showed that the long time limit of the correlation function
%\begin{equation}
%f_D^q = \lim_{t \rightarrow \infty} \phi_q(t),
%\end{equation}
%called the ergodicity breaking parameter,
%causes a bifurcation at the transition temperature.
%Also, in this letter, we use this value to judge the glass transition.
%Though, strictly speaking, 
%we should take $q$ as a characteristic length scale of the glass,
%this scale has not been obvious.
%Here, we especially take $q\rightarrow\infty$,
%so that we measure one point correlation function.
%\begin{equation}
%f_D^{\infty} = \lim_{t \rightarrow \infty} \sum_i \langle n_i(0) n_i(t) \rangle.
%\end{equation}
The precise determination of the transition point is difficult,
because the relaxation becomes very slow near the transition point.
However, fortunately a 
precise determination is not necessary for the thermodynamic study
we wish to carry out.
For this purpose, it is only important that
our binary lattice gas model
has a concentration glass transition.
(We have found that this model has another type of transition, 
a glass transition as a function of temperature.
However, this transition is not involved in the present work.)

Now we investigate the thermodynamics of the model.
For this purpose, it is convenient to slightly
modify the model described above.
First, we change the boundary conditions.
Among the four boundaries $i_x=0, i_y=0, i_x=L+1$ and $i_y=L+1$,
we consider the pair of boundaries $i_y=0$ and $i_y=L+1$
to be still periodic, while
we consider the other pair $i_x=0$ and $i_x=L+1$ to be
perfectly impermeable.
Next, we introduce a potential which represents a movable piston
into the model.
The piston potential, denoted by $V_a(i)$,
is added to the original, "bare" Hamiltonian ${\cal H}_0$,
so that the full Hamiltonian ${\cal H}$ is given by
\begin{equation}
{\cal H} = {\cal H}_0 + \sum_i V_a (i) ( n_i^A + n_i^B ).
\end{equation}
We represent the piston as a quadric potential:
\begin{equation}
V_a(i) = \begin{cases}
K \, (i_x-a)^2  & (i_x>a) \cr 
0 & (i_x\le a). \cr
\end{cases}
\end{equation}
By changing the value of $a$,
we carry out the operation of compressing or expanding the system,
and we refer to its value as the "piston position".
In our numerical simulations,
$a$ was changed as follows.
The piston position is initially $L$, and
at this time, the system is in the gas state.
Then we compress the system gradually until $a$ is $L_0$.
It is then gradually returned to $L$.
We used the following simple form of $a(t)$:
\begin{equation}
a(t)=\begin{cases}
-v t + L & (0\le t\le T) \cr
v (t -T)+ L_0 & (T \le t \le 2T), \cr
\end{cases}
\label{protocol}
\end{equation}
where $v$ is a parameter that represents the speed of the operation
and $T$ is defined by $T=(L-L_0)/v$.
When we make $L_0$ sufficiently small,
the system experiences the glass transition during the compression.
With the transition between the gas and glass states caused in this manner,
we investigate the thermodynamic properties of the glass state.
The quantity that we measure in the simulation
is the work performed on the system, defined by
\begin{equation}
W[a(t)]= \int \frac{\partial V_a}{\partial a} ( n_i^A + n_i^B ) \, da.
\end{equation}
We are particularly interested in the quasistatic limit of this function,
and we write this limiting form as $W_{qs}$.
In equilibrium,
$W_{qs}$ is equal to the Helmholtz free energy difference
between the initial and final configurations.
%\begin{equation}
%W_{qs}[a(t)] = F(a(t))-F(a(0)),
%\end{equation}
%where $F$ denotes the equilibrium free energy of the system.
However, because a glass state is non-equilibrium in nature,
it is not certain if $W_{qs}$ also can be used to 
construct the free energy in this case.
(Here, we note that we use the word "quasistatic" to mean simply
that the process is carried out sufficiently slowly
compared to the $\beta$-relaxation time scale.)
The main question we wish to answer in this letter is whether
$W_{qs}$ is a useful thermodynamic function in the description of the glass state.
For this purpose, we investigate the $v$ dependence of the work
$W[a]$ in a complete cycle of compression and expansion
defined by (\ref{protocol}).
If $\lim_{v \rightarrow 0} W[a] = 0$ in this cycle,
$W_{qs}$ constitutes a thermodynamic function of the glass state,
as it is in equilibrium thermodynamics.

\begin{figure}[h]
\begin{center}
\epsfile{file=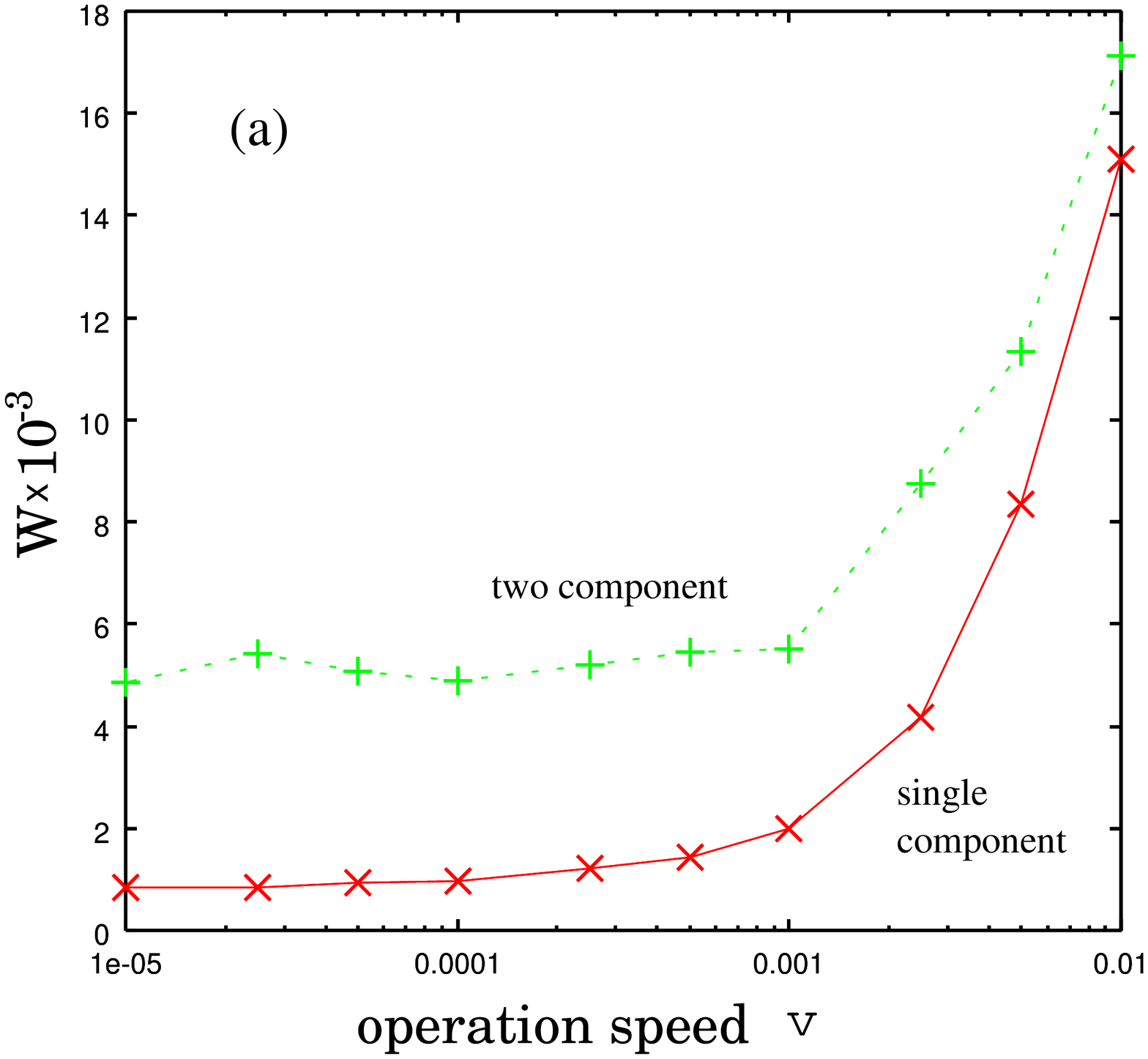,scale=0.38}
\epsfile{file=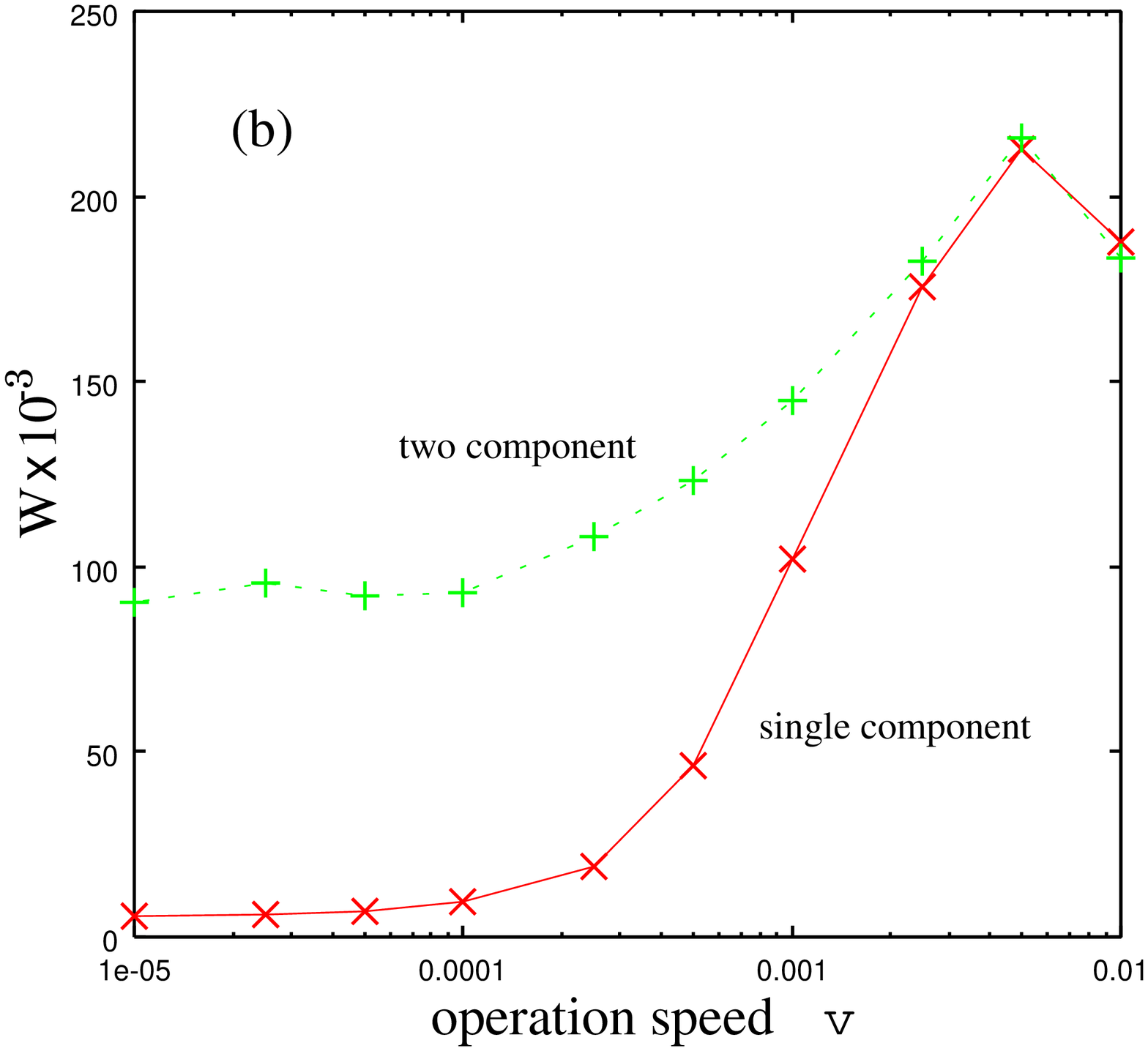,scale=0.38}
\end{center}
\caption{The irreversible work as a function of the speed of
the compression-expansion operation for both lattice gases.
The speed of operation $v$ is
measured in units of the inverse of the time needed to change the system length
by the inter-site distance.
(a) The parameter values are here $L=15$, $T=1000$, and 
the particle numbers are $(N^A,N^B)=(60,0)$
and $(N^A,N^B)=(30,30)$.
For each case,
the amount of work converges to a finite value
in the quasistatic limit.
The amount of quasistatic irreversible work in the two component case 
is much larger
than that in the single component case.
(b) Here the parameter values are $L=30$, $T=1000$, and
the particle numbers are $(N^A,N^B)=(240,0)$ and $(N^A,N^B)=(120,120)$.
We see a more distinct difference between the two component and single component
cases here than in (a).
}
\end{figure}
In this simulation we set
the parameters as $L=15$, $T=1000$ and $N^A,N^B = 30$.
$L_0$ is taken sufficiently small
as the system becomes a dense packing state at $a=L_0$.
With the parameter values given above,
the glass transition occurs at approximately $a = 9$.
In Fig. 2 (a) we display simulation results for $W[a]$ as a function of $v$.
For comparison, in the same figure,
we also plot the results obtained using a
Fuchizaki-Kawasaki type system with $N^A=60, N^B=0$.
As we see, in this case the system always remains in the gas state.
We can clearly see the effect of the glass transition
from this comparison. 
For each case,
we see that the work converges to a finite value
in the quasistatic limit,
but we see that the irreversible work performed in the two component 
case is much larger than that in the single component case.
In the single component case,
the amount of irreversible work ins very small,
and it appears to be due to a lattice effect.
This is demonstrated by the fact that the
amount of this work does not change when we set
the values of coupling constants to zero.
Thus we presume a slight amount of the work of the single component
case to be almost zero.
Contrastingly,
in the two component case,
irreversible work is essential
to the glass transition.
This implies that there is no thermodynamic function on
characterizing the glass state.

%Inneglegible irreversible work is, by its definition,
%considered to be based on the stress anomaly.
%Therefore let us focus on the internal stress field
%of the binary lattice gas.
%We define the stress of the model by
%\begin{equation}
%p_{i}^{\alpha}=(n_{i_{\alpha}+1}-n_{i_{\alpha}-1})n_i,
%\end{equation}
%and we show the stress field of the most compressed state $a=L_0$
%in Fig. 6.
%We can see the chain-like stress structure there.

\begin{figure}[h]
\begin{center}
\epsfile{file=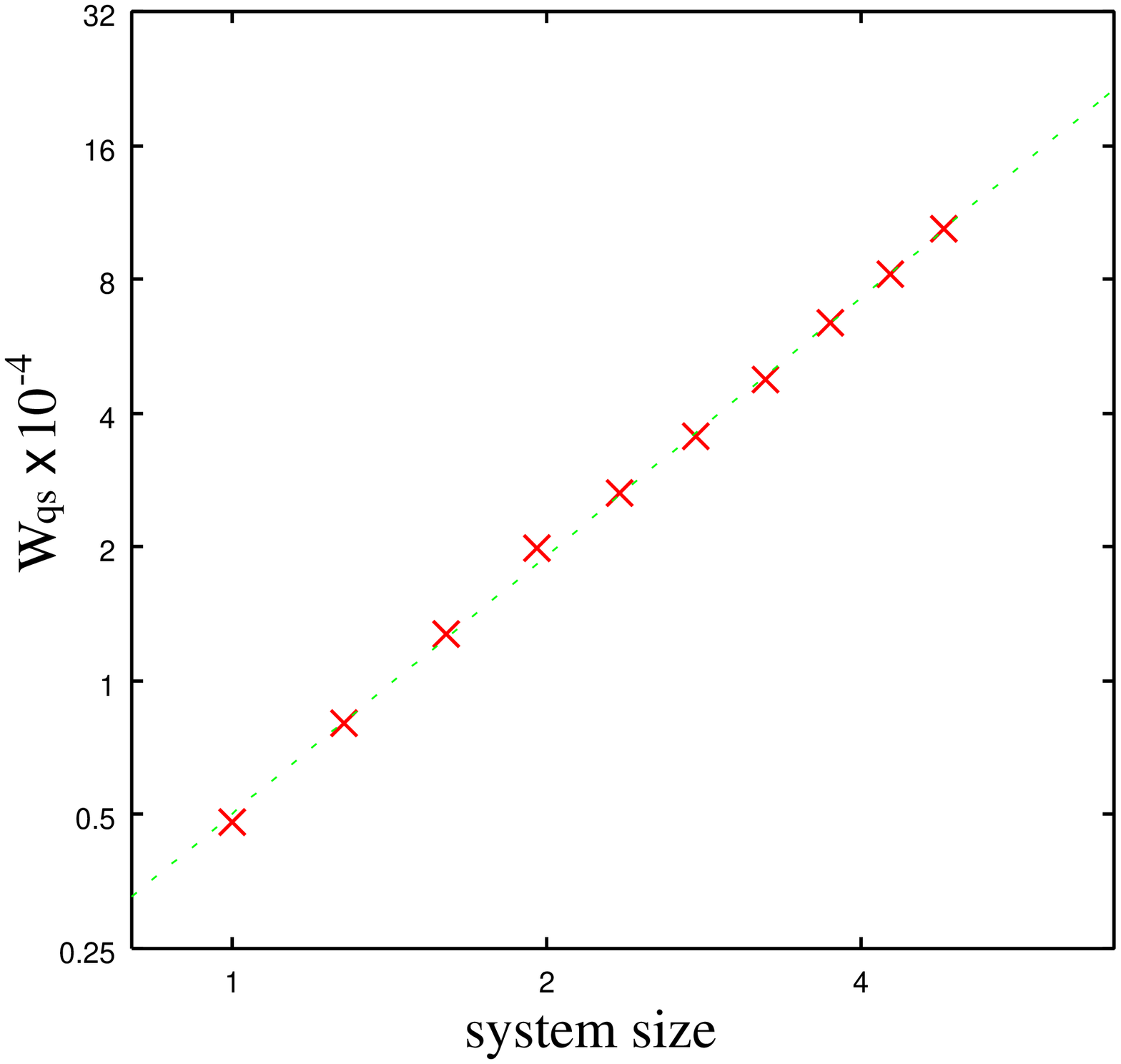,scale=0.4}
\end{center}
\caption{Log-log plot of the system size dependence of 
the quasistatic irreversible work in
the two component case.
The density is the same in each case.
System size is normalized as $15 \times 15$ is equal to 1.
The dashed line represents the fitting function $W \sim ($system size$)^2$.}
\end{figure}
The effect of the glass transition is seen more explicitly 
when we consider the dependence of the quasistatic irreversible work
on the system size.
In Fig.2 (b) we display the simulation results for $W[a]$ as a function of $v$ 
for two systems with four times more lattice points, but the same density:
$L=30$ and $(N^A,N^B)=(120,120)$ for the 
two component case,
and $L=30$ and $(N^A,N^B)=(240,0)$
for the single component case.
We find larger difference here between 
the value of $W_{qs}$ for the one component and two component systems
than in the $L=15$ case.
In Fig.3 we display the system size dependence of $W_{qs}$ for the two component system,
with equal densities in each case.
We find that the the irreversible work exhibits an anomalous and nonextensive
system size dependence: $W_{qs} \sim (L \times L)^2$.
The reason for this dependence is not yet clear.

From these results,
we conclude that 
the glass state does not conform to conventional thermodynamics laws.
We cannot define a thermodynamic function describing  the glass state, and
we need a theory which deals with the inevitable irreversibility directly. 
We point out here that approximately a decade ago,
Langer and Senthna reported that
the glass transition exhibits
a hysteresis of the entropy~\cite{LANG}.
Our results may be considered
the free energy version of their results.

We have arrived at an interesting conclusion.
From the operational point of view,
the glass transition can be described as
a generation of irreversibility, even in the quasistatic limit.
For the description of a glass,
it is thus not possible to use a simple extension of conventional thermodynamics.
%By the way, we already know a famous inevitable irreversibility
%``kinetic friction''.
%One of the Amonton-Coulomb low,
%which is the old law of friction, indicates that
%the kinetic frictional force is finite 
%even at quasistatic movement~\cite{PER}.
It is interesting that the macroscopic nature of our lattice gas
may be represented by the transition between a Newtonian fluid and
a Bingham fluid~\cite{FERRY},
because 
the inevitable generation of irreversibility in our system
can be regarded as a generation of yield stress.
In fact, we can determine the yield stress from Fig.2.
However, at present we do not know of a clear relation 
between a glass and a Bingham fluid.
The most important properties of glasses
are disorder and metastability, and
it is possible that these properties correspond to
properties of Bingham fluids,
but this correspondence is yet to be found.
In any case,
it may be productive to recognize that 
the glass transition may be one of the route to Bigham nature. 
%At present, the most received model of friction 
%is Frenkel-Kontrova model, however,
%there is no common theory of the source of friction.
%We consider that the theory describing macroscopic natures of the glass
%should also treat the friction explicitly,
%and expect an appearance of a nonequilibrium theory
%which unifies many glass theories and friction in the future.

\acknowledgments

The author would like to acknowledge K. Kaneko, T. Ikegami, S. Sasa
and other menbers of their community for valuable comments.
The author also acknowledges M. Kawasaki, K. Sato and K. Sekimoto
for useful suggestions, and ING and UGW for support with my reserch.

\end{document}